\documentclass[conference]{IEEEtran}
\IEEEoverridecommandlockouts
\usepackage{amsmath,amssymb,amsfonts}
\usepackage{dsfont}
\usepackage{algorithm}
\usepackage{algorithmic}
\usepackage{graphicx}
\usepackage{textcomp}
\usepackage{xcolor}
\usepackage{pgf}
\usepackage{tikz}
\usetikzlibrary{arrows,automata}
\usepackage{verbatim}
\usepackage{subcaption}
\usepackage{url}
\usepackage{balance}
\usepackage{cite}

\begin{document}

\title{Autonomous Maintenance in IoT Networks via AoI-driven Deep Reinforcement Learning}

\author{
	\IEEEauthorblockN{George Stamatakis\IEEEauthorrefmark{1},
		Nikolaos Pappas\IEEEauthorrefmark{2},
		Alexandros Fragkiadakis\IEEEauthorrefmark{1},
		Apostolos Traganitis\IEEEauthorrefmark{1}
		\IEEEauthorblockA{\IEEEauthorrefmark{1} Institute of Computer Science, Foundation for Research and Technology - Hellas (FORTH)}
		\IEEEauthorblockA{\IEEEauthorrefmark{2} Department of Science and Technology, Link\"{o}ping University, Campus Norrk\"{o}ping, Sweden}
		E-mails: \{gstam, alfrag, tragani\}@ics.forth.gr, nikolaos.pappas@liu.se}
}

\maketitle

\begin{abstract}
Internet of Things (IoT) with its growing number of deployed devices and applications raises significant challenges for network maintenance procedures. In this work, we formulate a problem of autonomous maintenance in IoT networks as a Partially Observable Markov Decision Process. Subsequently, we utilize Deep Reinforcement Learning algorithms (DRL) to train agents that decide if a maintenance procedure is in order or not and, in the former case, the proper type of maintenance needed. To avoid wasting the scarce resources of IoT networks we utilize the Age of Information (AoI) metric as a reward signal for the training of the smart agents. AoI captures the freshness of the sensory data which are transmitted by the IoT sensors as part of their normal service provision. Numerical results indicate that AoI integrates enough information about the past and present states of the system to be successfully used in the training of smart agents for the autonomous maintenance of the network.
\end{abstract}

\section{Introduction}
\label{sec:introduction}
The emergence of a massive Internet of Things (IoT) ecosystem poses new challenges for network maintenance processes. 
IoT networks are characterized by software, hardware, and communication protocols' diversity. 
Furthermore, they are typically comprised of a large number of devices which are often deployed in remote and harsh environments.
In this context the development of autonomous and intelligent maintenance procedures is deemed necessary in order to avoid the escalation of labor costs and the risk of outages.
Faults in IoT networks can be categorized into two distinct classes based on whether their mitigation requires physical intervention or not.
Autonomous maintenance targets the latter class of faults which can usually be mitigated by remote maintenance of the IoT devices. 
Examples of such remote maintenance procedures include the rebooting the IoT device's operating system, the update of a device's firmware or an application's software, restoring configurations and network engineering.

A large number of fault detection algorithms have been proposed in the past~\cite{muhammed2017analysis, zhang2018survey}. 
Currently, the majority of existing fault detection algorithms that utilize machine learning assume a centralized architecture and capitalize on the abundant computational resources of the cloud, yet they require the exchange of amounts of data related to the operational status of the IoT network.
This means that a large portion of the scarce bandwidth and energy resources of an IoT network are spent for transmitting maintenance related information instead of service provision.
Furthermore, the data necessary to make informed maintenance decisions often arrive late at the agent so that the timely maintenance of the network is difficult.

In this work, we consider an IoT application scenario whereby a number of IoT devices transmit sensory data to a monitor that resides in a cloud platform.
In such a scenario the sensory data often don't reach their destination due to malfunctioning IoT devices or network outages.
Our objective is to create an autonomous agent that would be capable to decide whether the IoT system is operational or not and, in the latter case, to be able to decide on the proper maintenance procedure, i.e., to perform maintenance of the sensors or of the network. In our study, we utilize the Age of Information (AoI) to achieve our goal, which quantifies freshness of information \cite{NOW-AoI, sunmodiano2019age, yates2020age}. To the best of our knowledge, there are currently no works to address the problem of autonomous maintenance in IoT networks with a similar approach to ours.

\subsection{Contributions}
The first contribution of this work is the formulation of the autonomous maintenance problem within the framework of Partially Observable Markov Decision Processes (POMDPs) and the
utilization of Deep Reinforcement Learning (DRL) techniques in order to train an agent for the task.
The second contribution of this work is that the autonomous agent makes decisions based on the freshness of the sensory data received at the monitor and thus does not load the network with extra traffic.
We utilize the AoI metric in order to model the freshness of the sensory data at the monitor and, despite the partial observability of the IoT network's true operational status, we derived numerical results which indicate that AoI integrates enough information about the current and the past states of the system to be successfully used as a reward signal for the training of autonomous agents.
Finally, since current DRL techniques require a large number of interactions between the learning agent and the environment, we assume that a simulation model of the IoT network is available.

\section{Related work}
A large number of fault and anomaly detection algorithms have been proposed for wireless sensor networks~\cite{muhammed2017analysis, zhang2018survey, mahapatro2013fault, jurdak2011wireless}. 
The majority of these algorithms utilize a variety of algorithmic and machine learning techniques in order to establish a model of the system's normal behavior.
Subsequently, the algorithm receives data related to the system's input and uses the learned model to calculate the system's expected output. 
The deviation between the calculated output values and those measured from the system is what guides the algorithm's decisions.
This process is typically iterative and involves human experts' knowledge in order to establish what is normal operation and what is not.
Considering the complexity of IoT networks we aimed at developing an agent that learns efficient maintenance procedures without the prior establishment of a system's model.
This is further justified by the fact that in many cases repairing a fault and restoring the system's operation can be easily done with generic procedures, e.g. by rebooting, by updating software or by restoring configurations, whereas finding the root cause of a fault is often a difficult, expensive and time consuming process. 

Although there exists a number of works that address the problem of fault detection and maintenance from a decision theoretic point of view, to the best of our knowledge none of them considers IoT systems as their reference system or follows an approach similar to our work.
More specifically, the authors in~\cite{cois1996fault} consider a network application scenario and develop an algorithm that detects and isolates faults by deciding the optimal dynamic sequence of diagnostic tests. 
In~\cite{verma2002probabilistic} the authors utilize a POMDP framework for the detection and mitigation of faults in robots. 
Their solution approach is based on the representation of the state space with a belief state vector and the utilization of particle filters for the estimation of their values.
In~\cite{langseth2003decision} the authors present a troubleshooting process based on Bayesian Networks which is optimal in terms of the number of troubleshooting steps. 
The troubleshooting steps are performed by humans and refer to electromechanical equipment such as printers. 
In~\cite{rish2005adaptive} the authors deal with the problem of real time fault detection and diagnosis in large distributed computer systems.
Under the assumption that the system is in a faulty state they develop an algorithm that finds the smallest set of probes with the property that at least one probe will fail under any type of fault.
In~\cite{zhong2019deep} the authors consider the problem of anomaly detection in a system comprised of an integer number of independent processes. 
The operational status of each process is monitored, with some degree of accuracy, by a sensor which sends this information to the decision algorithm. 
Furthermore, the authors assume that the decision algorithm can only communicate with a single sensor at a time and thus it is important to establish a policy that selects the sensor to be queried at each stage so that the time to locate the anomalous process is minimized. To find this policy the authors utilize DRL. 
In~\cite{gsnptaAlarms} the authors utilized AoI in order to train autonomous agents that are capable of reserving scarce energy resources in order to monitor processes and devices which exhibit faulty behavior.
Finally, in~\cite{lopez2020application} the authors train agents to detect intrusions in a network. They assume the existence of a data set with entries labeled as ``normal'' or ``intrusion'' and sample the data set in order to generate artificial trajectories of system states so that they can train smart agents with DRL algorithms.

Other works that consider AoI in IoT setups can be found in \cite{Corneo2019,GuIoT2019,Abd-Elmagid2018, DongIoT2020, StamatakisIoT2020,BeyturICNC2020}. 

\section{System Model}
\label{sec:systemModel}
\begin{figure}[h]
	\centering
	\includegraphics[scale=0.8]{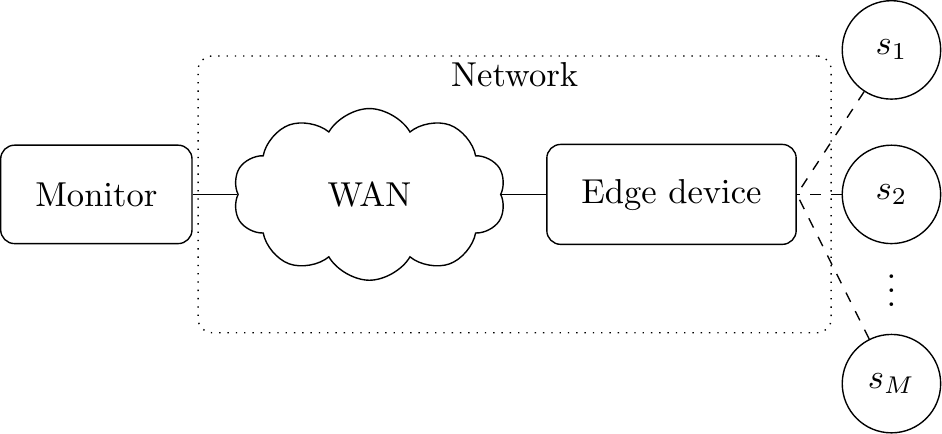}
	\caption{Basic IoT setup.}
	\label{fig:systemDiagram}	
\end{figure}

We consider the system presented in Figure~\ref{fig:systemDiagram}. It is comprised of $M$ sensors that transmit status updates to an edge device over a wireless link.
Subsequently, the edge device forwards the status updates to a cloud based monitoring application via a Wide Area Network (WAN).
We assume that time is slotted and indexed by $t \in \mathbb{Z}^+$.

Each sensor is modeled as an independent two state time-homogeneous Markov process. 
Let $s_t^i \in \{0,1\}$ be the state of the $i$-th sensor's Markov process at the beginning of the $t$-th time slot. 
When $s_t^i$ has a value of 0 (1), the $i$-th sensor's operational status is healthy (faulty).
We assume that each sensor will remain in the same state for the duration of a time-slot and, afterwards, it will make a probabilistic transition to another state as dictated by the state transition probability matrix $P_s$.
Furthermore, at the beginning of each time slot each sensor will generate a status update with probability $P_0^i$, when in a healthy state, and with probability $P_1^i$, when in a faulty state. 
We assume that $P_0^i > P_1^i$.
All sensors that generate a status update packet will try to transmit it to the edge device. 

Similarly, we model the operational status of the network, which includes the wireless link, the edge device and the WAN as its subsystems, as a two state time-homogeneous Markov process.
Let $s_t^n \in \{0,1\}$ denote the state of the network's Markov process at the beginning of the $t$-th time slot. 
When $s_t^n$ takes a value of 0 (1) the operation of the network is healthy (faulty). 
We assume that the network will remain in the same state for the duration of a single time-slot and, subsequently, it will make a transition to another state as dictated by the transition probability matrix $P_n$. 
When in a healthy state, the network will forward successfully a status update to the monitoring application with probability $P_0^n$, whereas a faulty network will succeed in forwarding the status update with probability $P_1^n$. 
We assume that $P_0^n > P_1^n$.

In this work we consider the problem of an autonomous agent, that resides at the monitor and has to learn how to perform maintenance on the IoT network described above.
The agent doesn't have access to the true operational status of the system, yet it has access to information related to the freshness of the status updates it receives from the sensors.
To characterize the freshness of the status updates received at the monitor we utilize the recently introduced AoI metric. 
AoI was defined in \cite{kaul2012real} as the time that has elapsed since the generation of the last status update that has been successfully decoded by the destination, i.e., 
\begin{equation}
\label{eq:classicalAoI}
    \Delta(t) = t - U(t)
\end{equation}
where $U(t)$ is the time-stamp of the last packet received at the destination at time $t$.
We use $\Delta_t^i,\ t = 0, 1, \dots$, to denote the AoI of the $i$-th sensor at time $t$.
At the beginning of each time-slot the agent will consider the value of the AoI metric for all sensors and subsequently, it will take one of three actions, i. ``No-maintenance'' ii. ``Network-maintenance'' and iii. ``Sensors-maintenance''. 
From these three actions only the ``No-maintenance'' action is cost free.
If at the beginning of a time slot the agent decides on the ``Network-maintenance'' action and the network is in a faulty state then it will be brought back to a healthy state.
Similarly, if the Sensors maintenance action is selected then all sensors that are in a faulty state will be brought back to a healthy state at the next time slot.
As mentioned above, all maintenance actions induce a cost on the system and thus the agent should try to avoid taking maintenance actions when the corresponding system units are in a healthy state.

\section{Problem Formulation}\label{sec:problemFormulation}
In this section we formulate the decision problem presented above as a Partially Observable Markov Decision Process (POMDP). 
The analysis presented in this section will facilitate the understanding of the system's operation and, subsequently, the interpretation of the numerical results.
\subsubsection*{State Space}
At the beginning of the $t$-th time-slot the state of the system is represented by the column vector,
\begin{equation}
	s_t = [s_t^1, s_t^2, \dots, s_t^M, s_t^n]^T
\end{equation}
where, as described in Section~\ref{sec:systemModel},  $s_t^i \in\{0, 1\}$ and $s_t^n \in \{0,1\}$ indicate that the operational status of the $i$-th sensor and the operational status of the network respectively and $T$ is the transpose operator. We denote with $\mathcal{S}$ the set of available states and emphasize that the true state of the system is unknown to the agent at time $t$.

\subsubsection*{Actions}
The set of actions available to the agent is denoted with $\mathcal{A} = \{0, 1, 2\}$, where $0, 1, \text{and } 2$ represent the "No-maintenace", "Network-maintenance" and "Sensors-maintenance" actions respectively. 
All three actions are available in all system states. 
Finally, we denote the action taken by the agent at the beginning of the $t$-th time slot with $a_t \in \mathcal{A}$. 

\subsubsection*{Random variables}
Given the current state of the system and the action taken by the sensor the system will make a stochastic transition to a new state.
The transition will occur at the end of the $t$-th time slot and it is determined by the realization of the following idependent random variables,
$W_h^{i,t} \in \{0,1\}$ and $W_h^{n,t} \in \{0,1\}$ which represent, for the $i$-th sensor and the network, respectively, the random transition to a new health status by the end of the $t$-th time slot.

The probability distributions for $W_h^{i,t}$ and $W_h^{n,t}$ are determined by the state of the system and the action taken by the agent at the beginning of the $t$-th time slot. 
More specifically, in case the "No-maintenance" and "Network-maintenance" actions are selected, i.e., $a_t = 0 \text{ or } 1$, we have, $\text{Pr}\{W_h^{i,t}=l\} = P_s^{kl}$, where $P_s^{kl}$ is the element of the transition probability matrix $P_s$ that corresponds to a transition from health status $k \in \{0,1\}$ at the $t$-th time slot to health status $l \in \{0,1\}$ at $(t+1)$. 
However, if "Sensors-maintenance" action is selected, i.e., $a_t = 2$,  then $W_h^{i,t}$ will be equal to 0 (healthy state) with probability $1$.

The probability distribution for $W_h^{n,t}$ is defined in a similar way, i.e., in case the "No-maintenance" and "Sensors-maintenance" actions are selected, we have $\text{Pr}\{W_h^{n,t} = l\} = P_n^{kl}$, where $P_n$ is the transition probability matrix for the network's Markov process and,
in case the "Network-maintenance" action is selected, we have $W_h^{n,t} = 0$ with probability $1$.

\subsubsection*{System Dynamics}\label{sec:systemDynamics}
The system's state at the beginning of the (${t+1}$)-th time slot is determined by the realization of random variables $W_h^{i,t}$ and $W_h^{n,t}$, which, as presented above, depend on $s_t$ and $a_t$.
More specifically we have, $s_{t+1}^i = W_h^{i,t}$ and $s_{t+1}^{n,t} = W_h^{n,t}$ which indicates that the system has the Markov property.

\subsubsection*{Observations}
At the beginning of the $t$-th time slot the agent observes the value of the AoI metric for each sensor. 
Let $\Delta_t = [\Delta_t^1, \Delta_t^2, \dots, \Delta_t^M]^T$ be the vector whose elements are the individual AoI values of the sensors.

The evolution over time of the AoI value for the $i$-th sensor depends on the following \emph{independent} random variables,
\begin{itemize}
	\item $W_{g}^{i, t} \in \{0, 1\}, i=1,\dots, M$, which represents the random event of a status update generation by the $i$-th sensor.
	\item $W_{n}^{i, t }\in \{0, 1\}$, which represent the random event of the network's successful ($W_n^{i,t} = 1$) or unsuccessful ($W_n^{i,t}~=~0$) delivery of the $i$-th sensor's status update from the sensor to the monitor.
\end{itemize}
The conditional probability distribution for random variable $W_{g}^{i, t}$ is, 
$\text{Pr}\{W_g^{i,t} = 0 | s_t^i = h\} = P_h^i$ and $\text{Pr}\{W_g^{i,t} = 1 | s_t^i = h\} = 1- P_h^i$ where $h \in \{0,1\}$.
Similarly, for $W_{n}^{i, t }$ we have $\text{Pr}\{W_n^{i,t} = 0 | s_t^n = h\} = P_h^n$ and 
$\text{Pr}\{W_n^{i,t} = 1 | s_t^n = h\} = 1- P_h^n$ where $h \in \{0,1\}$.

The evolution of AoI over time for each sensor is given by,
\begin{equation}
	\Delta_{t+1}^i = 
	\begin{cases}
		1, &	\text{if } W_g^{i, t} = 1 \text{ and }  W_n^{i,t} = 1 \\
		\min\{\Delta_{\max}, \Delta_t + 1\}, & \text{if } W_g^{i,t} = 0  \text{ or } W_n^{i,t} = 0  
	\end{cases}
\end{equation}
where $\Delta_{max}$ indicates an excessive level of staleness for the status update information beyond which there is no point in further increasing the AoI value.
The probability distributions of $W_g^{i, t}$ and $W_n^{i,t}$ depend on the network's and the sensors' health status at the previous time slot. 
This attribute, along with the Markov Property of the system's dynamics, indicate that the problem of optimally selecting actions based on AoI observations constitutes a Partially Observable Markov Decision Process (POMDP).

\subsubsection*{Transition reward function}
By the end of each time slot, the agent receives a reward which is given by
\begin{equation}\label{eq:transitionReward}
r_t = \frac{1}{\psi_c \cdot \sum_{j\in \mathcal{A}} \mathds{1}_{a_t\in\{j\}}\cdot c_j + \psi_{\Delta}\cdot\bar{\Delta}_t},
\end{equation}
where, $\mathds{1}_{a_t\in\{j\}}$ is the indicator function which assumes a value of $1$ when action $a_t$ equals $j \in \mathcal{A}$ and is zero otherwise. 
Weights $\psi_c$ and $\psi_{\Delta}$ are used to form a weighted sum of the maintenance cost $c_j$ associated with action $j$ and the average $\bar{\Delta}_t$ calculated over all $\Delta_{t}^i$, i.e., $\bar{\Delta}_t = \frac{1}{M}\sum_{i=1}^{M} \Delta_t^i$.
In this work we assume that the maintenance cost associated with the ``No-maintenance'' action is zero, i.e., $c_0 = 0$.
Equation~(\ref{eq:transitionReward}) assumes its maximum value when the ``No- maintenance'' action is selected and the AoI equals $1$ for all sensors, i.e., information at the monitor is as fresh as possible. 
Whenever the agent selects a maintenance action or the sensors fail to send a status update to the monitor due to unaddressed faults the reward diminishes. 
Finally, equation~(\ref{eq:transitionReward}) indicates that $r_t$ is a function of all random variables presented in this section, i.e., it depends on the random vector $W_t = [W_h^{i,t}, W_h^{n,t}, W_{g}^{i, t}, W_{n}^{i, t }]$ but not on its previous values. 

\subsubsection*{Total reward function}
In a POMDP the agent doesn't have access to the actual state of the system, thus, to optimally select actions it must utilize all previous observations and actions up to time $t$~\cite[Chapter 4]{B17}.
Let $h_t = [\Delta_0, \Delta_1, \dots, \Delta_{t}, a_0, a_1, \dots, a_{t-1}]$ be the \emph{history} of all previous observations and actions, with $h_0 = \{\Delta_0\}$. 
Furthermore, let $\mathcal{H}$ be the set of all possible histories for the system at hand.

The agent must find a policy $\pi^*$ that maps each history in $\mathcal{H}$ to a probability distribution over actions, i.e., $\pi:~\mathcal{H} \rightarrow P(\mathcal{A})$, so that the expected value of the total reward accumulated over a infinite horizon is maximized. 
Let $\Pi$ be the set of all feasible policies for the system at hand, then, assuming that the agent's policy is $\pi \in \Pi$ and has an initial history $h_t$ the expected value of the total reward accumulated over a infinite horizon is,
\begin{equation}\label{eq:ValueFunction}
	V_{\pi}(h_0) = \mathop {\mathbb{E}}_{W_0, W_{1}, \cdots} \Big[ \sum_{t=0}^{\infty} \gamma^t r_t \Big\vert h_0, \pi \Big],
\end{equation}
where expectation $\mathbb{E}\{\cdot\}$ is taken with respect to the joint distribution of the random variables in $W_t$ for $t= 0, 1,\dots$ and the given policy $\pi$. Finally, $\gamma$ is a discount factor that reduces the importance of future rewards.
Our objective is to find the optimal policy $\pi^*$ which is defined as $\pi^* = \arg\underset{\pi \in \Pi}{\max}\, V_{\pi}(h_0)$.

Besides the state value function presented in Equation~(\ref{eq:ValueFunction}), a function that has a central role in deriving optimal policies is the optimal state-action value function $Q^*(h,a)$ which represents the maximum value of the expected cumulative reward achievable by selecting action $a$ given the history vector is $h$ at time $t$ and subsequently following the optimal policy over an infinite horizon, i.e., 
\begin{equation}\label{eq:StateActionValueFunction}
	Q^*(h, a) = \max_{\pi \in \Pi} \mathop {\mathbb{E}}_{W_k, W_{k+1}, \dots} \Big[ \sum_{k=t}^{\infty} \gamma^k r_k \Big\vert h_t = h, a_t = a, \pi \Big].
\end{equation}
$Q^*(h,a)$ is also called the $Q$-value function and satisfies the Bellman equation~\cite{B17},
\begin{equation}\label{eq:Bellman}
Q^*(h,a) = \mathbb{E} [ r + \gamma \max_{a' \in A} Q^*(h', a')]
\end{equation}
where $h_t = h, a_t = a, h_{t+1} = h'$ and $a_{t+1} = a'$ which is at the basis of all algorithms that derive optimal $Q$-value functions. 
If the action space is comprised of a small number of discrete actions and $Q^*(h, a)$ is known, selecting the optimal action $a^*$ for any history $h$, is straightforward, 
\begin{equation}
	a^* = \arg \max_{a\in \mathcal{A}} Q^*(h,a).
\end{equation}

\section{Learning approximately optimal policies}\label{sec:optimalPolicy}
Finding an optimal policy for a POMDP typically involves the reduction~\cite{silver2010monte}\cite[Chapter 4]{B17} of the original problem to a Markov Decision Process (MDP) whereby the agent has perfect knowledge of the system's state.
As described in the previous section, this reduction process involves, among other things, the introduction of a new state space that is either represented by the history vector $h_t$ or by the belief state vector~\cite[Chapter 4]{B17} whose elements represent the conditional probability of the system's actual state $s_t$ given $h_t$, i.e., $b(s,h) = \text{Pr}\{s_t = s \vert h_t = h\}$.
In both cases the resulting MDP is typically intractable~\cite{papadimitriou1987complexity} or undecidable~\cite{madani1999undecidability}. 
A notable exception to this result occurs when the optimal policy possesses certain structural properties~\cite{krishnamurthy2016partially}.
However, a model of the system is required in order to derive the structural properties of the optimal policy, if such properties exist. 
Assuming the existence of an analytical model is too restrictive an assumption for the problem of autonomous maintenance. 

To deal with the state space explosion resulting from the ever increasing dimension of $h_t$ over time we resort to approximations.
More specifically, we approximate $h_t$ with the most recent vector of AoI values, i.e., $\Delta_{t}$.
However, even under this approximation, it is inefficient to represent $Q(\Delta,a)$ as a table and proceed with learning a different $Q(\Delta,a)$ value for each state-action pair.
This is because $\Delta_t$, belongs to a vector space that has $\Delta_{\max}^M$ possible elements since $\Delta_t = [\Delta_t^1, \Delta_t^2, \dots, \Delta_t^M]$ and $\Delta_{t}^i \in \{0, 1, \dots, \Delta_{\max}\}$. 
To address this problem we utilize DRL algorithms. 
In DRL $Q^*(\Delta,a)$ is approximated by an Artificial Neural Network (ANN) $Q(\Delta,a;\theta)$ where $\theta$ denotes the weights and biases of the ANN.
DRL algorithms capitalize on the ability of ANNs to generalize to previously unseen state-action pairs.
What is more, a DRL agent learns policies through interaction with the system or a simulated model of the system, thus an analytical model is not required.
Recent research on DRL has lead to prominent successes, such as~\cite{mnih2015human} and~\cite{silver2017mastering}, that have spurred the interest for the application of DRL algorithms in the context of autonomous communications and networking systems~\cite{lei2020deep}.
In this work we develop two algorithms for the maintenance scheduling problem. The first one is based on the Deep Q-Network (DQN)~\cite{mnih2015human} algorithm and the second one is based on the Advantage Actor-Critic (A2C) algorithm \cite{sutton2018reinforcement, mnih2016asynchronous}.
Next we present the adaptation of these algorithms to our problem.

\subsection{M-DQN}
The M-DQN algorithm is based on the DQN algorithm~\cite{mnih2015human}. 
An episodic version of the M-DQN algorithm is presented in Algorithm~\ref{alg:MDQN}. 
In order to deal with past limitations of ANN-fitted $Q$-value functions~\cite{riedmiller2005neural} DQN introduced two important techniques.
The first one was that $Q^*(\Delta, a)$ was approximated by two ANNs instead of one.
The two ANNs are called the online and the target network and are denoted with $Q(\Delta,a;\theta)$ and $Q(\Delta,a;\theta^-)$ respectively.
Both ANNs accept as input the vector $\Delta_t$ and return the corresponding $Q$-values for all possible actions. 
The parameters $\theta$ and $\theta^-$ of the two ANNs are initialized to the same values.
The parameters of the online network are updated after each iteration while those of the target network are assigned periodically the corresponding values of the online network ($\theta$) and are kept fixed throughout the remaining period of $T_c$ time slots.
The second technique used by the DQN algorithm is that it stores past transitions of the form $(\Delta_t, a_t, r_t, \Delta_{t+1})$ in an experience buffer with capacity $C_D$.

At each time slot $t$ the algorithm receives $\Delta_t$ and selects an action $a_t$. 
Action selection is done with two different methods. 
The fist one is the classic $\epsilon$-greedy policy which selects $a_t$ either uniformly from $\mathcal{A}$ or as the action with the maximum $Q$-value estimate. 
The first case occurs with probability $\epsilon$ while the second case occurs with probability $1 - \epsilon$.
The second method is a biased $\epsilon$-greedy (b$\epsilon$g) policy, which we found to significantly reduce the number of episodes required for the agent to learn an efficient policy. 
An agent following the b$\epsilon$g-policy selects with probability $\epsilon$ the "No-maintenance" action and with probability $1 - \epsilon$ it follows the classic $\epsilon$-greedy policy.
The b$\epsilon$g-policy is efficient due to two reasons. 
The first reason is that it incorporates our prior belief that the ``No-maintenance" action will be used most of the time in a typical IoT network. 
The second reason is DRL algorithms typically start with a high $\epsilon$ probability which results in a frequent selection of the two maintenance actions.
As a result the system is frequently brought to a fully operational status depriving the agent from the chance to experience states with high $\Delta_t^i$ values. 
The M-DQN algorithm that utilizes the biased $\epsilon$-greedy policy is presented in Algorithm~\ref{alg:MDQN} as M-b$\epsilon$g-DQN. 
Finaly, the selected action is provided to the model simulator which returns a reward $r_t$ and the next observation $\Delta_{t+1}$.

Transition $(\Delta_t, a_t, r_t, \Delta_{t+1})$ is stored in $D$. 
Subsequently the algorithm selects a random batch of transitions from $D$ in order to train the online network.
Unlike consecutive transitions, which are highly correlated, the randomly selected transitions from $D$ are independent and this is a prerequisite for the effective use of stochastic gradient descent.
For the $i$-th transition of the sampled batch DQN uses $r_i$ and the target network to compute $Y_i = r_{i+1} + \gamma \max_{a_{i+1}} Q(\Delta_{i+1}, a_{i+1};\theta^-)$ 
which is an estimate of the right hand side of the Bellman Equation~(\ref{eq:Bellman}) and serves as the target value for the online network.
Subsequently, DQN uses the online network to calculate the approximation $Q(\Delta_i,a_i;\theta)$ of the target value $Y_i$.
The training of the online network is done with gradient descent and the objective is to minimize the loss $\mathcal{L}_i$, which is defined to be the squared distance between $Y_i$ and $Q(\Delta_i,a_i;\theta)$.
It becomes evident that by keeping $\theta^-$ constant for a period of time DQN succeeds in keeping the target values fixed as well. 
Thus the training process of the online network resembles more the typical training process of an ANN whereby the targets (labels) remain constant throughout the training process.

\begin{algorithm}
	\caption{M-DQN and M-b$\epsilon$g-DQN}
	\label{alg:MDQN}
\begin{algorithmic}[1]
	\STATE Initialize randomly the weights $\theta$ of $Q(s,a;\theta)$
	\STATE Set $\theta^- = \theta$ for $\hat{Q}(s,a;\theta^-)$
	\STATE Set $D$ to capacity $C_D$.
	\FOR {episode = $1$ to $E$} 
		\FOR {$t = 1$ to $T$}
			\STATE Observe $\Delta_t = [\Delta_t^0,\dots, \Delta_t^M]$
			\STATE Select action $a_t$ following the $epsilon$-greedy policy for M-DQN and the biased $\epsilon$-greedy policy for M-b$\epsilon$g-DQN.
			\STATE Execute action $a_t$ in the simulated model and collect reward $r_t$ and next observation $\Delta_{t+1}$
			\STATE Store transition ($\Delta_t$, $a_t$, $r_t$, $\Delta_{t+1}$) in $D$
			\STATE Sample a random mini-batch of $B$ transitions, $(\Delta_i, a_i, r_i, \Delta_{i+1}), i= 1,\dots, B$, from $D$
			\STATE Calculate target values, \begin{equation*}
						Y_i = \begin{cases} 
							r_i\text{, if episode terminates at step } i+1, \\
							r_i + \gamma \max_{a'} \hat{Q}(\Delta_{i+1}, a';\theta^-)\text{, otherwise}
							\end{cases} 
					\end{equation*}
			\STATE Calculate loss $\mathcal{L}_i = (Y_i-Q(\Delta_i,a_i;\theta))^2$
			\STATE Perform gradient descent on $\mathcal{L}_i$ with respect to $\theta$
			\STATE Every $T_c$ steps set $\theta^- = \theta$
		\ENDFOR
	\ENDFOR
\end{algorithmic}
\end{algorithm}

\subsection{M-A2C}
The M-A2C algorithm is an Advantage Actor-Critic (A2C) algorithm~\cite{sutton2018reinforcement, mnih2016asynchronous}.
An episodic version of the algorithm is presented in Algorithm~\ref{alg:MA2C}.
As is typical for A2C algorithms the value and the policy functions are approximated by two ANNs, denoted with $V(\Delta;\theta_v)$ and $\pi(\Delta;\theta)$, which are called the critic and the actor networks respectively.
The objective of the M-A2C algorithm is to find a parameter vector so that the $V(\Delta;\theta_v)$ is maximized.
To this end it utilizes a stochastic gradient ascent algorithm whereby $\theta$ is updated according to, 
\begin{equation}
\theta \gets \theta + \alpha \nabla_{\theta_v} V(\Delta;\theta_v).
\end{equation}
From the Policy Gradient theorem with a baseline~\cite[Chapter 13]{sutton2018reinforcement}, we have,
\begin{equation}
	\label{eq:PolicyGradientTheorem}
	\nabla_{\theta_v} V(\Delta;\theta_v) = \mathds{E}\bigg[ \big(\hat{Q}_{\pi}(\Delta_t, a_t) - V(\Delta_t;\theta_v)\big)  \nabla_{\theta} \ln \pi(a_t|\Delta_t; \theta)\bigg]
\end{equation}
where $\hat{Q}_{\pi}(\Delta_t, a_t)$ is an estimate of $Q_{\pi}(\Delta_t, a_t)$ and $V(\Delta_t;\theta_v)$ plays the role of a baseline against which $\hat{Q}_{\pi}(\Delta_t, a_t)$ is compared.
Let $t$ be the current time slot and $\rho$ a fixed period of time slots, then $T_{\rho} = t + \rho$ represents the time interval of $\rho$ time slots ahead of $t$.
Furthermore let $T_{\mathds{1}_e}$ be the index of the time slot where the current episode ends, i.e., it holds that $T_{\mathds{1}_e}>t$. 
We define $T_u = \min\{T_{\mathds{1}_e},T_{\rho}\}$ to be equal to the time interval or $\rho$ time slots ahead of current time $t$ unless the episode ends before $T_{\rho}$ in which case $T_u$ is the interval up to the end of the episode. $\hat{Q}_{\pi}(\Delta_t, a_t)$ is derived from sample trajectories generated according to policy $\pi$ that have a duration of $T_u$ time slots and is given by
\begin{equation}
	\label{eq:A2C_Qvalue_estimate}
\hat{Q}_{\pi}(\Delta_t, a_t) = \sum_{k=0}^{K-1} \gamma^k r_{t+k} + \gamma^{K} V(\Delta_{t+K};\theta_v),
\end{equation}
where  $K = T_u - t$.
In Algorithm~\ref{alg:MA2C} the generation of the sample trajectory appears in lines~\ref{alg2:trajectoryStart} to~\ref{alg2:trajectoryEnd} and the iterative evaluation of expressions~(\ref{eq:PolicyGradientTheorem}) and~(\ref{eq:A2C_Qvalue_estimate}) appear in lines~\ref{alg2:PolicyGradientStart} to~\ref{alg2:PolicyGradientEnd}. 
Finally, the critic's loss function is,
\begin{equation}
	\mathcal{L}^c = [\hat{Q}_{\pi}(\Delta_t, a_t) - V(\Delta_k;\theta_v)]^2,
\end{equation}
where $\hat{Q}_{\pi}(\Delta_t, a_t)$ is given by expression~(\ref{eq:A2C_Qvalue_estimate}) and its iterative evaluation appears in line~\ref{alg2:critic}.

\begin{algorithm}
	\caption{M-A2C}
	\label{alg:MA2C}
	\begin{algorithmic}[1]
		\STATE Initialize randomly the weights $\theta_v$ of $V(\Delta;\theta_v)$
		\STATE Initialize randomly the weights $\theta$ for $\pi(\Delta;\theta)$
		\REPEAT
			\STATE Reset gradients $d\theta = 0$ and $d\theta_v = 0$
			\STATE $t_s = t$
			\STATE Get observation $\Delta_t$
			\REPEAT \label{alg2:trajectoryStart}
				\STATE Sample an action from policy $\pi(\Delta_t;\theta)$
				\STATE Receive reward $r_t$ and observation $\Delta_{t+1}$ from the simulated model
				\STATE $t \gets t + 1$
			\UNTIL {$t=T_u$} \label{alg2:trajectoryEnd}
			\STATE Set $R = \begin{cases} 0,\text{ If } s_t \text{ is terminal,} \\ V(\Delta_t; \theta_v), \text{otherwise}  \end{cases}$ \label{alg2:PolicyGradientStart}
			\FOR {$k$ in $t-1, t-2, \dots, t_s$}
				\STATE $R \gets r_i + \gamma R$
				\STATE Accumulate gradients:
				\STATE 	$d \theta \gets d\theta + (R-V(\Delta_k;\theta_v))\nabla_{\theta} \log \pi(a_k|\Delta_k; \theta) $
				\STATE 	$d\theta_v \gets d\theta_v + \frac{\partial (R-V(\Delta_k;\theta_v))^2}{\partial \theta_v}$ \label{alg2:critic}
			\ENDFOR \label{alg2:PolicyGradientEnd}
			\STATE Perform synchronous update of $\theta$ and $\theta_v$ using $d\theta$ and $d\theta_v$ respectively
		\UNTIL {$t > T$} 
	\end{algorithmic}
\end{algorithm}

\section{Results}
In this section, we evaluate numerically the M-DQN, M-b$\epsilon$g-DQN and M-A2C algorithms for a system comprised of four sensors.
Throughout all experiments, $\Delta_{\max}$ equals an episode's duration which was set to $5000$ time slots, the maintenance cost $c$ was set equal to $100$ for both the network and the sensors, $\psi_c$ was set to be equal to $\psi_{\Delta}$, and the discount factor $\gamma$ was set to $0.999$.

The ANN used for the online and target networks in the M-DQN and M-b$\epsilon$g-DQN algorithms is comprised of an input layer, whose nodes take as input the elements of $\Delta_t$ vector, followed by a fully connected linear transformation with 128 hidden nodes which are in turn followed by Rectified Linear Units (ReLU), and finally a fully connected output layer which has three nodes, one for each potential action. Furthermore, the M-DQN has an experience buffer with a capacity of $5\cdot 10^5$, the learning rate was set to $10^{-4}$, syncing between the online and target networks occurred every $2\cdot10^4$ time slots. $\epsilon$ was decreased from a value of $1$ to a value of $0.01$ over a period of $4\cdot 10^5$ time slots.

The ANN used for the actor of the M-A2C algorithm is comprised of an input layer, whose nodes take as input the elements of $\Delta_t$, followed by two fully connected layers with $32$ and $128$ hidden nodes respectively, and finally the output layer which is a fully connected layer with three nodes, one for each action. The actor shares its input and hidden layers with the critic which has an additional fully connected layer of 32 nodes, followed by a rectified nonlinearity and an output layer of a single node which is followed by a $\tanh$ nonlinearity.
In this work we used the Pytorch~\cite{NEURIPS2019_9015} deep learning library for the implementation of the aforementioned ANNs.

\begin{figure}[h]
	\centering
	\begin{subfigure}[]{0.49\columnwidth}
		\centering
		\includegraphics[width=\linewidth]{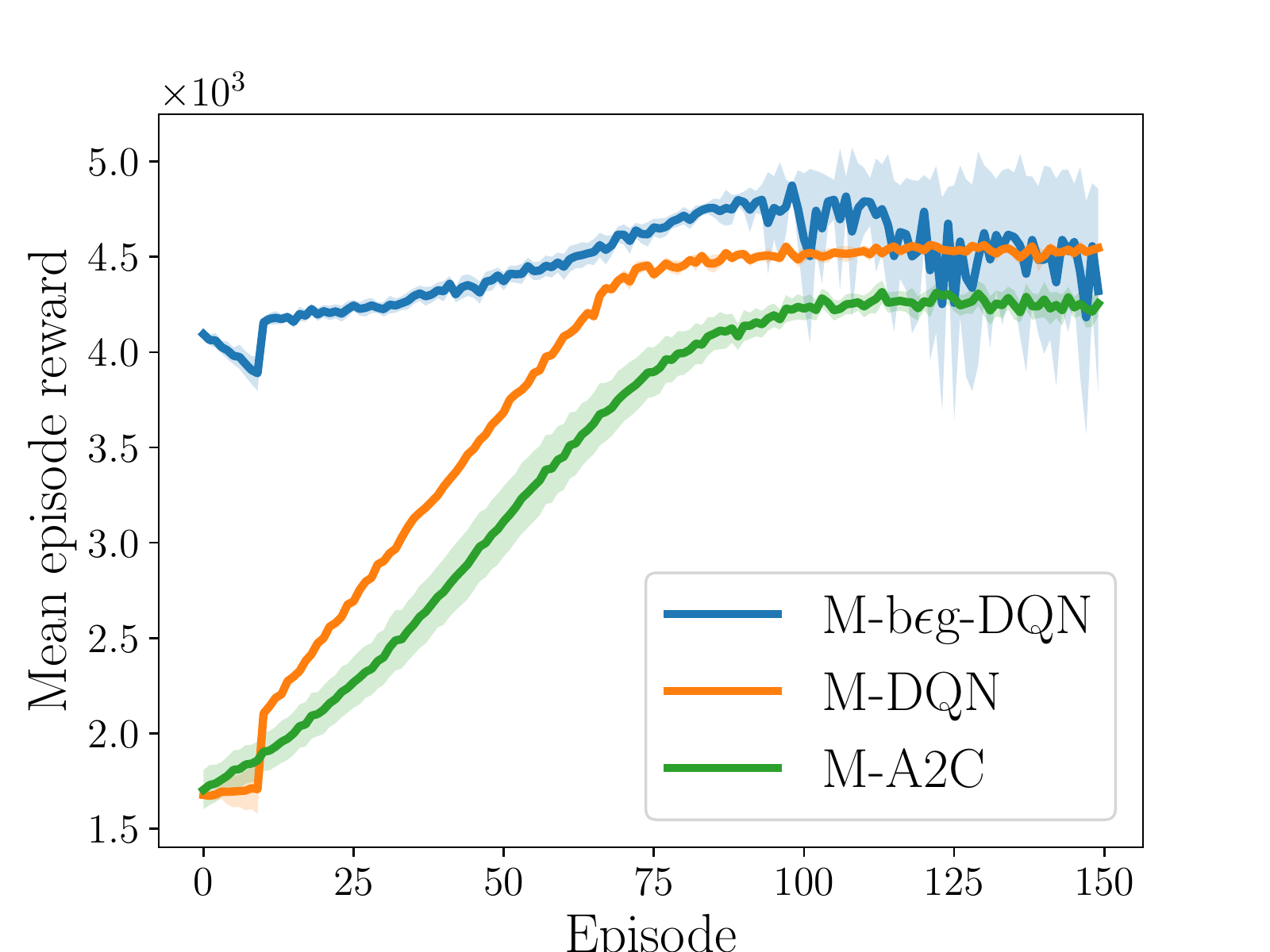}
		\caption{Permanent Faults}
		\label{fig:PermanentFautls}	
	\end{subfigure} 
	\begin{subfigure}[]{0.49\columnwidth}
		\centering
		\includegraphics[width=\linewidth]{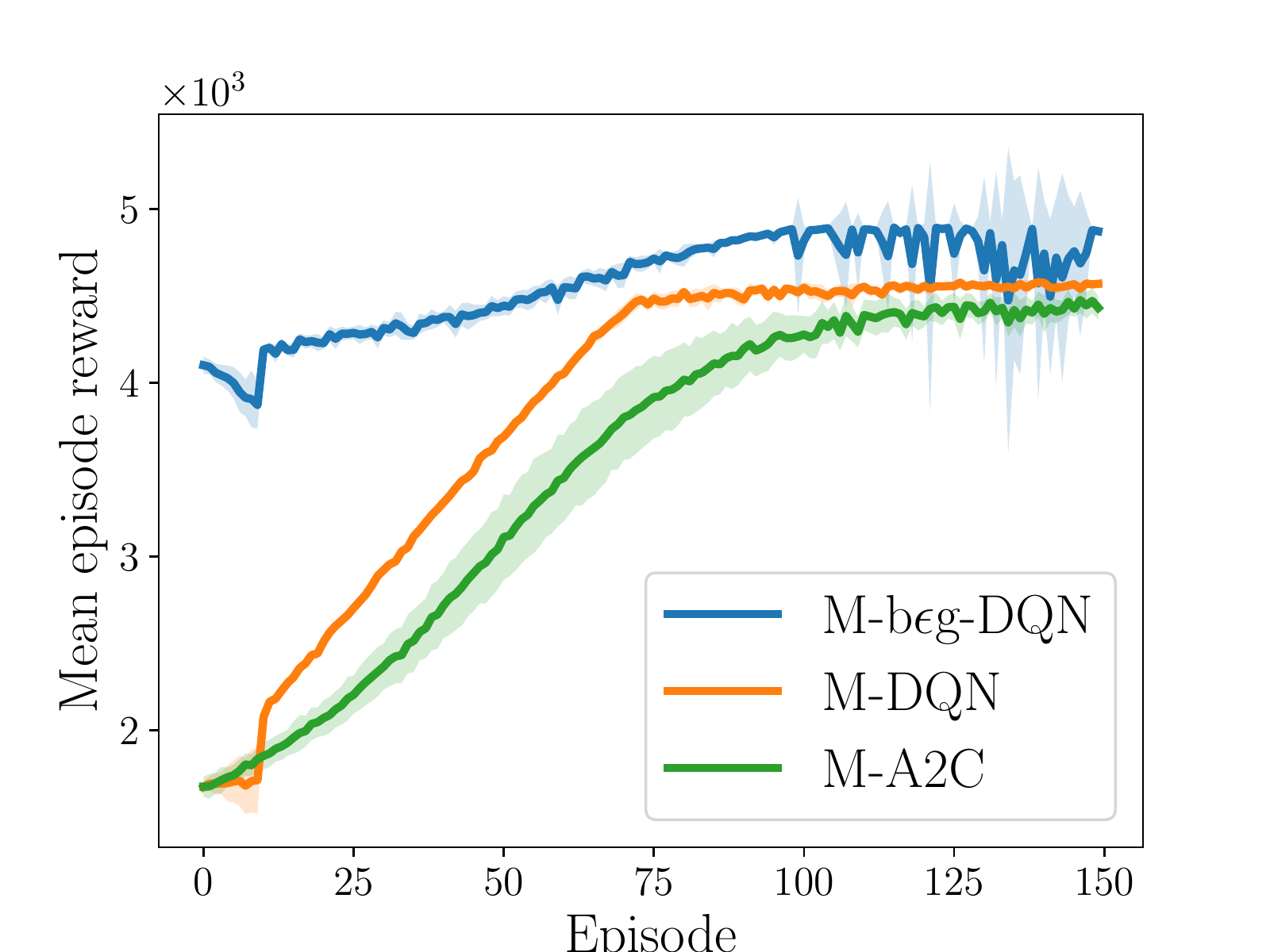}
		\caption{Intermittent Network Faults}
		\label{fig:IntermittentNetworkFaults}
	\end{subfigure} 
	\begin{subfigure}[]{0.49\columnwidth}
		\centering
		\includegraphics[width=\linewidth]{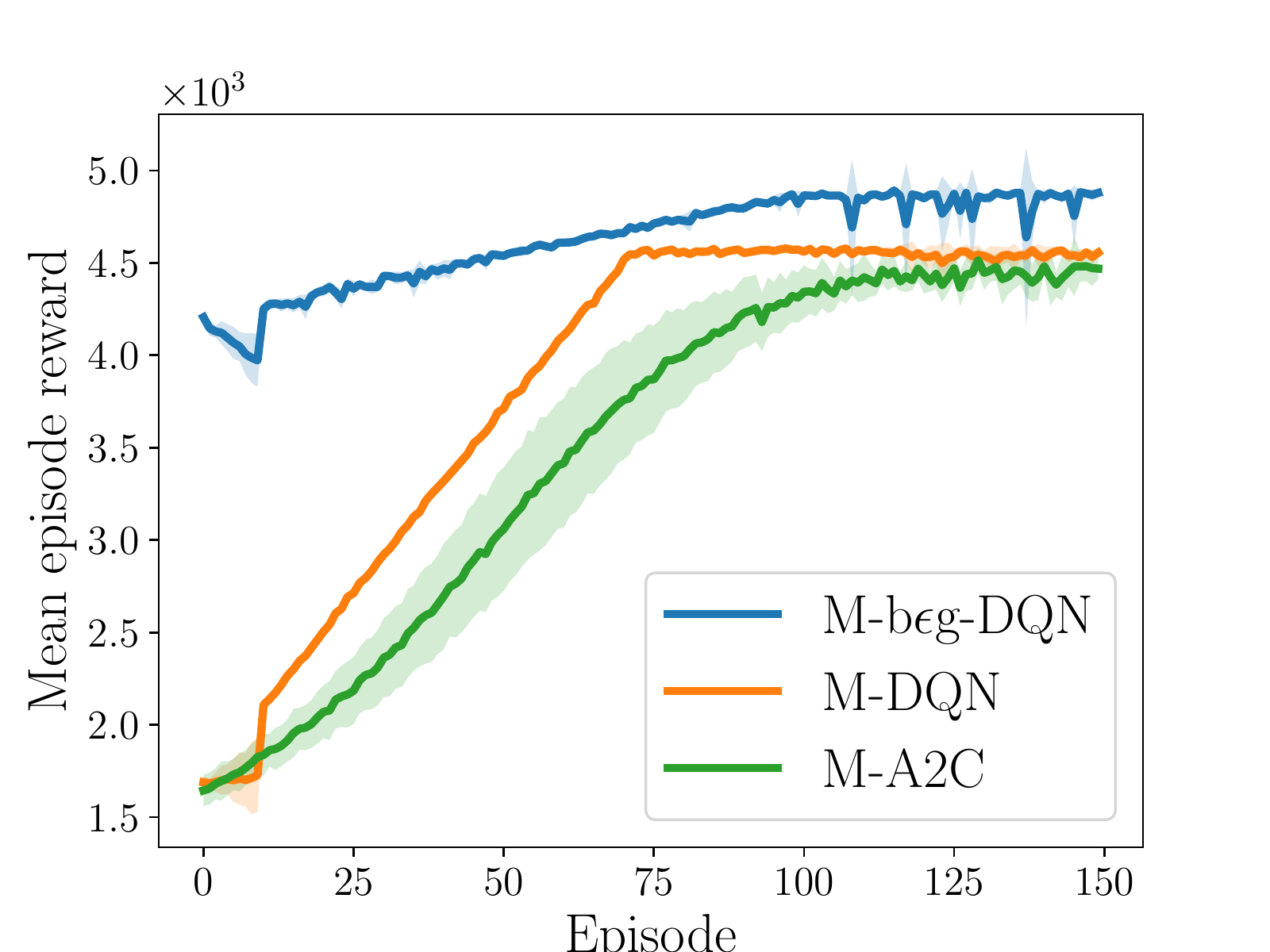}
		\caption{Intermittent Sensor Faults}
		\label{fig:IntermittentSensorFautls}
	\end{subfigure}
	\begin{subfigure}[]{0.49\columnwidth}
		\centering
		\includegraphics[width=\linewidth]{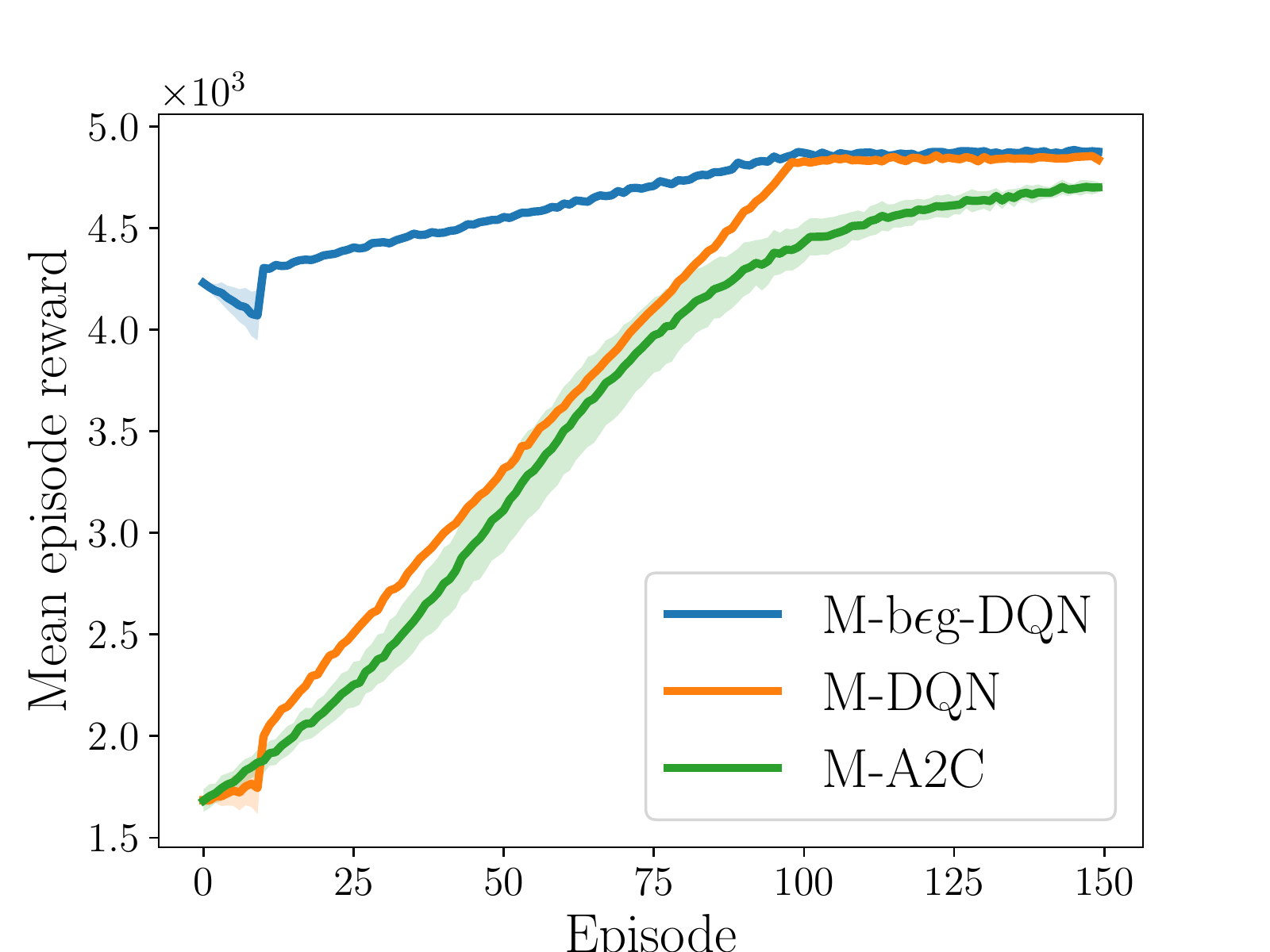}
		\caption{All Faults are Intermittent}
		\label{fig:AllFaultsIntermittent}
	\end{subfigure}\vspace{-0.05in}
	\caption{Mean episode reward with a $95$\% confidence interval calculated across $20$ training sessions each consisting of $150$ episodes.}
	\label{fig:MeanEpisodeRewardVsEpisode}
\end{figure}

In Figure~\ref{fig:MeanEpisodeRewardVsEpisode} we compare the ability of the M-b$\epsilon$g-DQN, M-DQN and M-A2C algorithms to learn efficient policies.
In all four scenarios of Figure~\ref{fig:MeanEpisodeRewardVsEpisode}, the probability that a sensor will generate a status update was set to $1$ ($0$) if the sensor was in a healthy (faulty) state, i.e., $P_0^s = 1$ ($P_1^s = 0$), and the probability that the network would forward a status update successfully was set to $0.99$ ($0$) in case the network was healthy (faulty), i.e., $P_0^n = 0.99$ ($P_1^n = 0$).
The four scenarios are differentiated by the network's and sensors' state transition matrices which are defined respectively as, $P_n = \left[ \begin{array}{cc}					
	P_n^{00} & 1-P_n^{00} \\
	1 - P_n^{11} & P_n^{11}
\end{array}\right]$ and 
$P_s = \left[ \begin{array}{cc}
		P_s^{00} & 1-P_s^{00} \\
		1 - P_s^{11} & P_s^{11}
			\end{array}\right]$.
In these definitions $P_n^{00}$ ($P_s^{00}$) represents the probability that the network (sensor) will make a transition to a healthy state given that it currently is in a healthy state, while $P_n^{11}$ ($P_s^{11}$) represents the probability that the network (sensor) will make a transition to a faulty state given that it currently is in a faulty state (also see Section~\ref{sec:problemFormulation}). In all four scenarios, $P_n^{00}=P_s^{00}=0.999$.
For the scenario with Permanent Faults presented in Figure~\ref{fig:PermanentFautls} all faults were permanent and thus $P_s^{11} = P_n^{11} = 1$. 
For the scenario with intermittent network faults, presented in Figure~\ref{fig:IntermittentNetworkFaults}, the only change from the first scenario was that $P_n^{11} = 0.9$. 
Similarly, for the scenario with intermittent sensor faults presented in Figure~\ref{fig:IntermittentSensorFautls} we set $P_s^{11} = 0.9$. 
Finally, in the scenario with intermittent network and sensor faults presented in Figure~\ref{fig:AllFaultsIntermittent},  $P_n^{11}=P_s^{11}=0.9$.
For each scenario of Figure~\ref{fig:MeanEpisodeRewardVsEpisode} and for each DRL algorithm in it we performed $20$ training sessions each consisting of a sequence of 150 episodes.
Let $R_{i,j}$ be the total reward accumulated over the duration of the $i$-th episode of the $j$-th training session, then the vertical axis for all scenarios of Figure~\ref{fig:MeanEpisodeRewardVsEpisode} presents $\bar{R}_i = \sum_j\frac{R_{i,j}}{20}$ along with its $95\%$ confidence interval. 
Considering the transition reward function presented in Expression~(\ref{eq:transitionReward}), which has a maximum value of $1$, the fact that the duration of the episode is $5000$ time slots, and assuming no failures, no failed transmissions, and no false positives on behalf of the agent, the maximum reward that can be accumulated is $5000$. The results of Figure~\ref{fig:MeanEpisodeRewardVsEpisode} exhibit that the policies learned by the three algorithms increase $\bar{R}_i$ towards the $5000$ limit albeit at different rates. M-DQN and M-b$\epsilon$g-DQN performed better compared to the M-A2C algorithm in all four scenarios. However, M-b$\epsilon$g-DQN learned good policies in less episodes than the M-DQN although its $\bar{R}_i$ values exhibit higher variance and especially for large values of $i$.

DRL algorithms learn policies with the objective to maximize the expected aggregate reward over a time horizon. 
On the other hand, in autonomous maintenance systems we are usually interested in the ability of the agent to detect and mitigate faults.
Thus, we have to verify that the reward function we chose for training agents results in a sane behavior as far as fault mitigation is concerned.
To this end, we trained four agents, one for each type of scenario presented in Figure~\ref{fig:MeanEpisodeRewardVsEpisode}, using the M-b$\epsilon$g-DQN algorithm. 
Subsequently, we assigned each agent the task of maintaining the system it was trained for, for a duration of $1000$ episodes.  

\begin{figure}[]
	\centering
	\includegraphics[scale=0.5]{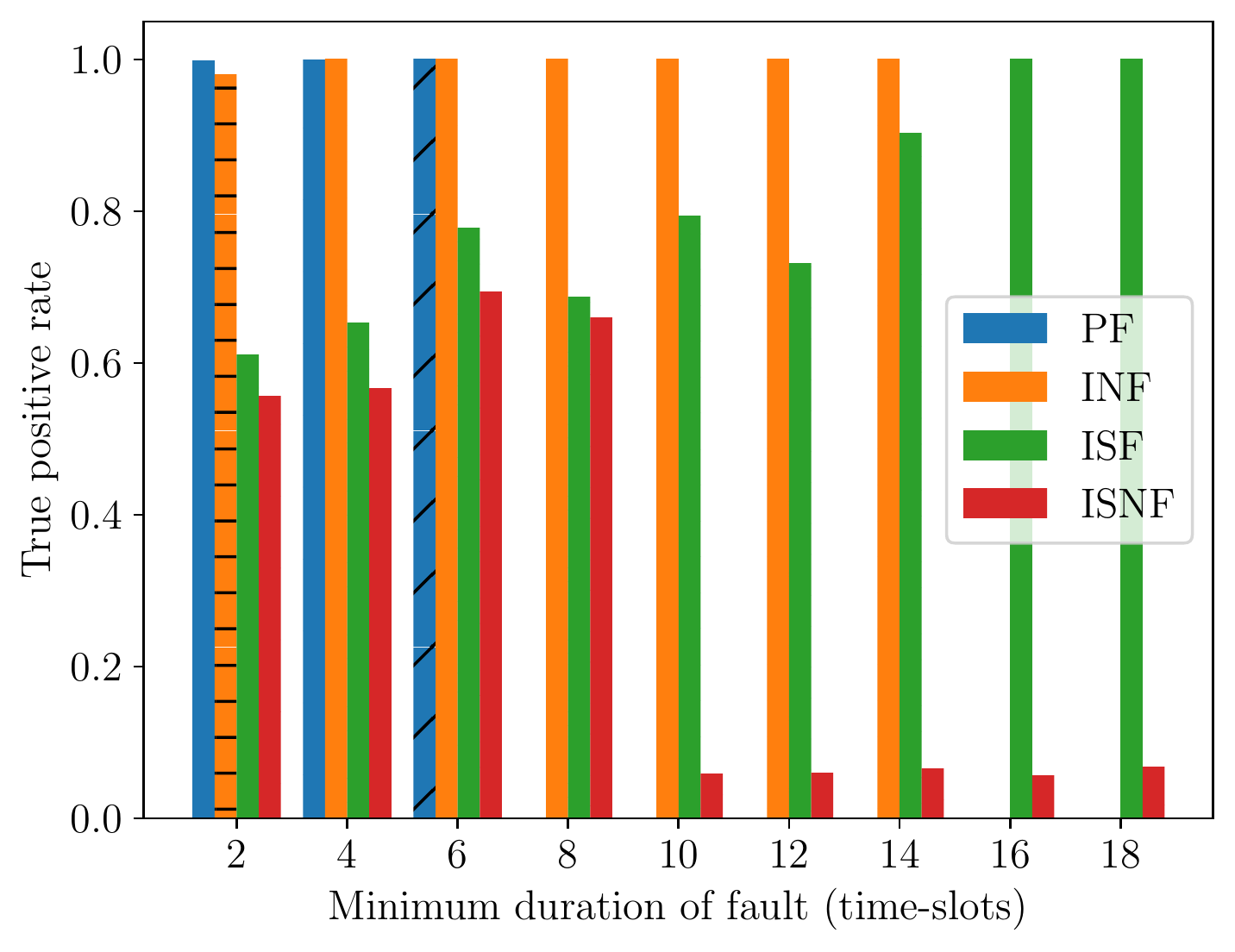}
	\caption{True positive rate for M-b$\epsilon$g-DQN agents tested in scenarios with Permanent Faults (PF), Intermittent Network Fautls (INF), Intermittent Sensor Faults (ISF) and Intermittent Network and Sensor Faults (ISNF).}
	\label{fig:truePositiveRate}
\end{figure}

Figure~\ref{fig:truePositiveRate} presents the true positive rate, i.e., the rate at which the corresponding agent detected existing faults in the system, versus the minimum duration of the fault measured in time slots. 
We note in Figure~\ref{fig:truePositiveRate} that for a system with permanent faults (PF in Figure~\ref{fig:truePositiveRate}) the agent trained for it was always able to identify an occurring fault. 
What is more, the lack of a blue colored bar for faults with duration larger than eight time slots indicates that all faults were detected and dealt with in less than eight time slots.
We observe a similar level of performance for the agent that was trained for a system with intermittent network faults (INF in Figure~\ref{fig:truePositiveRate}).
On the other hand, in the case of a system with intermittent sensor faults (ISF in Figure~\ref{fig:truePositiveRate}), the corresponding agent has a much lower true positive rate. 
This phenomenon can be attributed to two reasons.
The first reason is the reward function, defined in Equation~(\ref{eq:transitionReward}), whereby we use the mean AoI of all sensors in order to calculate the agent's reward. 
As a result, if a single sensor had failed, as was the case in our experiments, the reduction of the reward perceived by the agent was not significant enough to justify the cost of a maintenance action. 
However, as the duration of the unaddressed fault increased so did the AoI of the sensor and, eventually, the balance shifted towards taking the correct maintenance decision. 
This is indicated by the fact that all faults with a duration of $16$ time slots and more were identified correctly by the agent. 
The second reason, is the intermittent character of faults, i.e., there exist cases where the sensors returned to a healthy state by themselves and the fault was never identified by the agent. 
This is manifested in the cases of faults with a minimum duration of $8$ and $12$ time slots.
Finally, in the case of intermittent network and sensor faults (INSF in in Figure~\ref{fig:truePositiveRate}) we have, initially, a level of performance that resembles that for intermittent sensor faults.
However, for faults with a minimum duration that is larger than $8$ time slots the true positive rate of the agent diminishes.
This indicates that when both types of faults are intermittent and the duration of a fault exceeds a certain threshold the agent may not identify and mitigate the correct type of fault. 
To alleviate this problem one could utilize a larger history of past observations and actions at the cost of increasing exponentially the size of the state space.

\section{Conclusions}
In this work we address the problem of autonomous maintenance in IoT networks. We utilized DRL algorithms to train smart agents for the task and utilized the AoI metric as a reward signal for their training. Numerical results indicate that AoI integrates enough information about the past and present states of the system to be used successfully in training smart agents for the maintenance of the IoT network we consider in this work.

\section*{Acknowledgment}
This research has been financed by the European Union
and Greek national funds through the Operational Program
Competitiveness, Entrepreneurship and Innovation, under the
call RESEARCH – CREATE – INNOVATE (project code:
T1EDK-00070).

\ifCLASSOPTIONcaptionsoff
\newpage
\fi
\balance
\bibliographystyle{IEEEtran}
\bibliography{bibliography}

\end{document}